\documentclass{aa}
\usepackage{graphicx}
\usepackage{amssymb}
\usepackage{natbib}
\bibpunct{(}{)}{;}{a}{}{,}
\usepackage{txfonts}
\usepackage[dvips]{color}

\begin{document}
   \title{Chemical Self-Enrichment of H{\sc ii} Regions by the Wolf-Rayet Phase of an $85\mbox{M}_{\sun}$ star }

   \subtitle{}

   \author{D. Kr\"oger \inst{1}  \and G. Hensler \inst{2} \and T. Freyer \inst{1}}

   \institute{{Institut f\"ur Theoretische Physik und Astrophysik der Universit\"at 
         Kiel, D-24098 Kiel, Germany \\ 
         \email{danica@astrophysik.uni-kiel.de} } 
    \and {Institute of Astronomy, University of Vienna, T\"urkenschanzstr. 17, 
         A-1180 Vienna, Austria \\
         \email{hensler@astro.univie.ac.at} } }

   \offprints{D. Kr\"oger}

\date{Received  / Accepted }

\abstract{ It is clear from stellar evolution and from observations of
  WR stars that massive stars are releasing metal-enriched gas through
  their stellar winds in the Wolf-Rayet phase. Although H{\sc ii}
  region spectra serve as diagnostics to determine the present-day
  chemical composition of the interstellar medium, it is far from
  being understood to what extent the H{\sc ii} gas is already
  contaminated by chemically processed stellar wind.  Therefore, we
  analyzed our models of radiative and wind bubbles of an isolated
  $85\mbox{M}_{\sun}$ star with solar metallicity \citep{kfhy} with
  respect to the chemical enrichment of the circumstellar H{\sc ii}
  region.  Plausibly, the hot stellar wind bubble (SWB) is enriched
  with $^{14}\mbox{N}$ during the WN phase and even much higher with
  $^{12}\mbox{C}$ and $^{16}\mbox{O}$ during the WC phase of the star.
  During the short period that the $85\mbox{M}_{\sun}$ star spends in
  the WC stage enriched SWB material mixes with warm H{\sc ii} gas of
  solar abundances and thus enhances the metallicity in the H{\sc ii}
  region.  However, at the end of the stellar lifetime the mass ratios
  of the traced elements N and O in the warm ionized gas are
  insignificantly higher than solar, whereas an enrichment of $22\%$
  above solar is found for C.  Important issues from the presented
  study comprise a steeper radial gradient of C than O and a
  decreasing effect of self-enrichment for metal-poor galaxies.
   }

\titlerunning{Chemical Enrichment of H{\sc ii} Regions by WR stars}

\authorrunning{D. Kr\"oger et al.}

\maketitle
%
%________________________________________________________________

\section{Introduction}
\label{Introduction}

H{\sc ii} regions are considered as the most reliable targets for the
determination of the present-day abundances in the Interstellar Medium (ISM).
Nevertheless, heavy elements are released by massive stars into their 
surrounding H{\sc ii }~regions. The knowledge of its amount is of 
particular interest for the observation of very metal-poor galaxies. 
\citet{kunth} e.g. discussed the problem of determining the heavy-element 
abundance of very metal-poor blue compact dwarf galaxies from emission lines
of H{\sc ii} regions in the light of local self-enrichment by massive 
stars. This can happen during two
stages in their evolutionary course: In the Wolf-Rayet (WR) stage,
when the stellar wind has peeled off the outermost stellar layers, and
by supernovae of type~II (SNeII).
Because of their energetics, stellar winds and SNeII contribute their
gas to the hot phase. Therefore, it is of high relevance for observations
and the addressed question of the presented exploration to study the amount 
to what it can be mixed into the diagnosed H{\sc ii} regions.

As discussed by \citet{cm}, massive stars with an initial mass greater
than $25 \mbox{ M}_{\odot}$ evolve into WR stars with
enhanced mass-loss rates and chemically enriched stellar winds.

\citet{abbott} concluded from stellar evolution models that WR 
winds strongly contribute to the Galactic enrichment of $^{4}$He, 
$^{12}$C, $^{17}$O, and $^{22}$Ne, while contributing only moderately 
to the enrichment of $^{14}$N, $^{26}$Mg, $^{25}$Mg, and $^{16}$O.

For metallicities less than solar two effects reduce the heavy element
release by WR stars: First, the lower the
metallicity the more massive a star has to be to evolve through the WR
stages. Therefore, the number of WR stars decreases with decreasing
metallicity. \defcitealias{schaller:1992}{SSMM}\citet[][ hereafter
SSMM]{schaller:1992} found that with metallicity $Z=0.001$ the minimal initial
H-ZAMS mass for a WR star is $> 80 \mbox{ M}_{\odot}$. At $Z=0.02$ the
minimal initial mass is $> 25 \mbox{ M}_{\odot}$. Second, the lower
the metallicity the shorter are the WR lifetimes, and not all WR
stages are reached. At solar metallicity WR stars enter all three WR
stages (WNL, WNE, WC), whereas at $Z=0.001$ only the WNL phase is
reached \citepalias{schaller:1992}. The WR lifetime of an $85 \mbox{
  M}_{\odot}$ star, e.g., is $t_\mathrm{WR}= 0.204 \times 10^5 \mbox{ yr}$ at
$Z=0.001$ and $t_\mathrm{WR}= 4.008 \times 10^5 \mbox{ yr}$ at $Z=0.02$
\citepalias{schaller:1992}.

That WR stars play an important role for carbon enrichment of the ISM
at solar metallicity was shown by \citet{dtkl}. Their models predict
that the C enrichment by WR stars is at least comparable to that by
AGB stars while the enrichment of N is dominated by AGB stars and the
O enrichment is dominated by SNeII.
At other metallicities \citet{dray} found an increase in carbon
enrichment with increasing metallicity but a decrease in oxygen
enrichment. The N enrichment by WR stars is negligible compared to
that by other sources. Additionally, the total amount of N ejected by
a WR star generally decreases with decreasing metallicity.

\citet{hirschi:2004, hirschi:2005} showed that the consideration of stellar rotation
 in the Geneva stellar evolution code increases the yields
for heavy elements by a factor of $1.5 - 2.5$ for stars between $15
\mbox{ and } 30 \mbox{ M}_{\odot}$. For the more massive stars rotation
raises the yields of $^4\mbox{He}$ and other H-burning products like
$^{14}\mbox{N}$, but the yields of He-burning products like
$^{12}\mbox{C}$ are smaller. Additionally, for stars with $M \gtrsim
60 \mbox{ M}_{\odot}$, the evolution differs from that of non-rotating
stars by the following manner: Rotating stars enter the WR regime already 
in the course of their main-sequence.

As an extent to stellar evolution models mentioned above
hydrodynamical simulations of radiation and wind-driven H{\sc ii}
regions can trace the spatial distribution of the elements ejected by
the star, and, therefore, we can study the mixing of the hot
chemically enriched stellar wind with the warm photoionized gas.

%
%________________________________________________________________

\section{The model}
\label{model}

\defcitealias{tim}{Paper~I}
\defcitealias{timII}{Paper~II}
\defcitealias{kfhy}{Paper~III}

We analyze the model of an $85 \mbox{M}_{\sun}$ star in a series of
radiation and wind-driven H{\sc ii} regions around massive stars
\citep[ hereafter: Paper~I, Paper~II, Paper~III]{tim,timII, kfhy} 
%\citep[ Paper~III]{kfhy},
as described in \citetalias{kfhy}. A description of the numerical
method and further references are given in \citetalias{tim}.  The
time-dependent parameters of the $85 \mbox{ M}_{\sun}$ star with
``standard'' mass-loss and solar metallicity ($Z = 0.02$) during its
H-MS and its subsequent evolution are taken from
\citetalias{schaller:1992}, who used the relative abundance ratios of
\citet{anders} for the heavy elements.

An $85 \mbox{ M}_{\odot}$ star with solar abundance enters its WN stage
at an age of $t = 2.83 \mbox{ Myr}$.
The WR star enriches the combined SWB/H{\sc ii} region with $^{12}$C,
$^{14}$N, and $^{16}$O. In our 2D simulations we study the spatial
distribution of these tracer elements inside the SWB/H{\sc ii} region
as well as their time-dependent mass ratios as
$\mbox{M}_{^{12}\mbox{C}}/\mbox{M}_\mathrm{total}$,
$\mbox{M}_{^{14}\mbox{N}}/\mbox{M}_\mathrm{total}$, and
$\mbox{M}_{^{16}\mbox{O}}/\mbox{M}_\mathrm{total}$ in two different
temperature regimes: ``warm'' gas $\left(6.0 \times 10^3 \mbox{ K} \le
  T < 5.0 \times 10^4 \mbox{ K}\right)$ accounts for the gas of the
H{\sc ii} region, and ``hot'' gas, whose temperatures $\left( T \ge 5
  \times 10^4 \mbox{ K}\right)$ are reached inside the SWB.
Additionally, for the H{\sc ii} region diagnostic a degree of
ionization of $\ge 0.95$ must be reached.  The undisturbed ambient
medium outside the H{\sc ii} region as well as the material swept-up
by the expanding H{\sc ii} region are not counted since they are not
ionized. The observable abundances of H{\sc ii} regions are derived
for the warm phase only.

The mass fractions of $^{12}\mbox{C}$, $^{14}\mbox{N}$, and
$^{16}\mbox{O}$ in the circumstellar gas as well as of the star itself
are initially set to the solar values, taken from \citet{anders},
$4.466 \times 10^{-3}$, $1.397 \times 10^{-3}$, and $1.061 \times
10^{-2}$, respectively, in relation to H.
%
%________________________________________________________________

\section{Results}
\label{Results}

The spatial distribution of the tracer elements is shown as
$\log({\rho \left(\mathrm{element}\right)}/{\rho
  \left(\mathrm{element, solar} \right)})$ at different stages of
stellar evolution, where ``element'' means $^{12} {\mbox{C}}$
(Fig.~\ref{CNO}, middle column), $^{14} {\mbox{N}}$ (Fig.~\ref{CNO},
left column), or $^{16} {\mbox{O}}$ (Fig.~\ref{CNO}, right column),
respectively.  All plots cover the whole computational domain of $60
\mbox{ pc} \times 60 \mbox{ pc}$. The gas density is overlaid as a
contour plot.  Before the onset of the WN phase, these density ratios
are unity by definition.  With the onset of the WN phase the combined
SWB/H{\sc ii} region is chemically enriched by the WR wind.

We study the temporal evolution of $^{12}\mbox{C}$, $^{14}\mbox{N}$,
and $^{16}\mbox{O}$, respectively, normalized to solar values in the
two different temperature regimes of the combined SWB/H{\sc ii}
region: For the hot gas of the SWB the mass ratios are depicted in
Figure~\ref{C12_N14_O16_hot}. Figure~\ref{C12_N14_O16_warm} shows it
for the ``warm'' H{\sc ii} gas (see \S~\ref{model}).

\subsection{Chemical enrichment of the stellar wind bubble}
\begin{figure*}
\centering
\resizebox{18.5cm}{!}{\hspace{-3cm}
\includegraphics{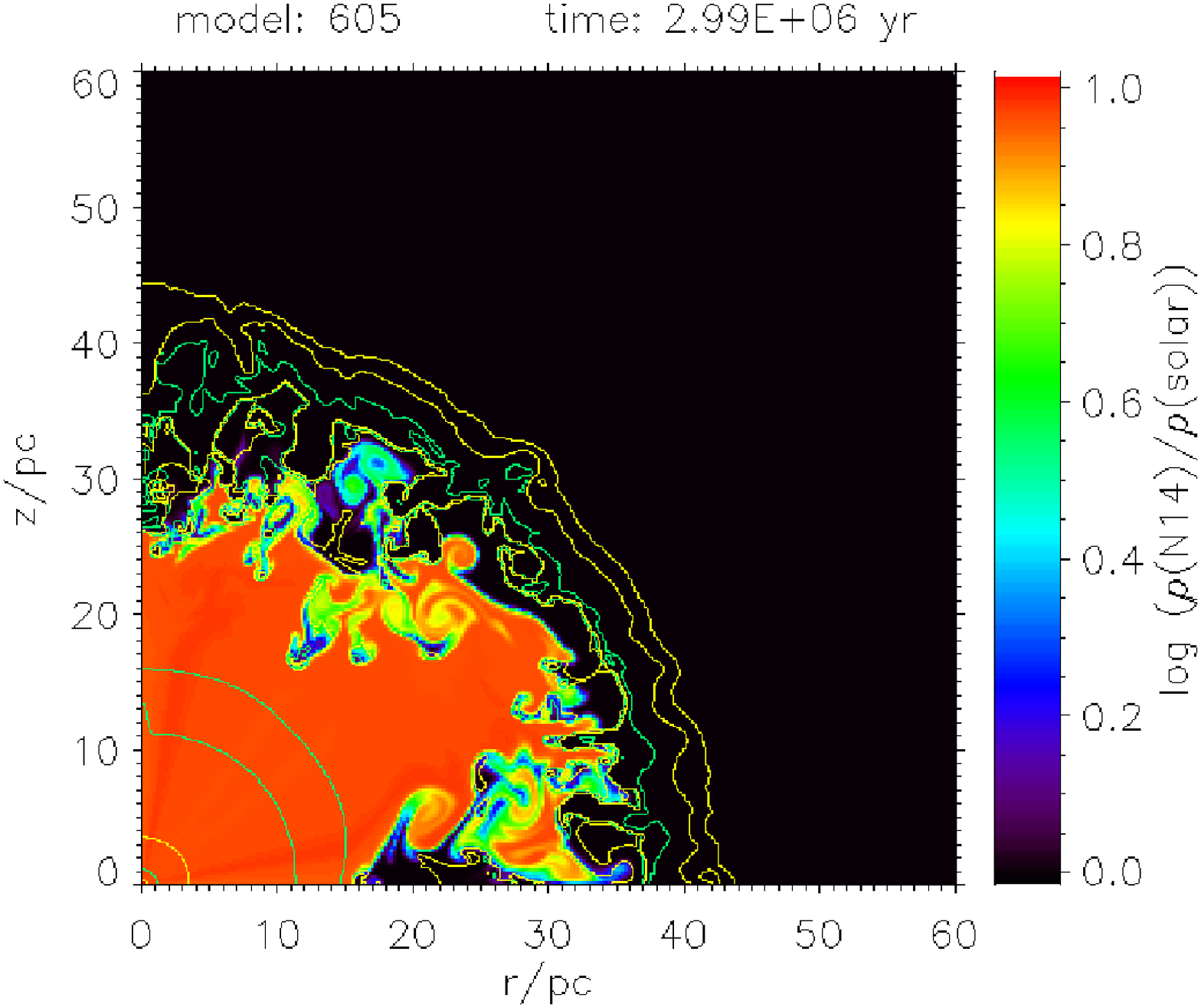}\hspace{-3cm}
\includegraphics{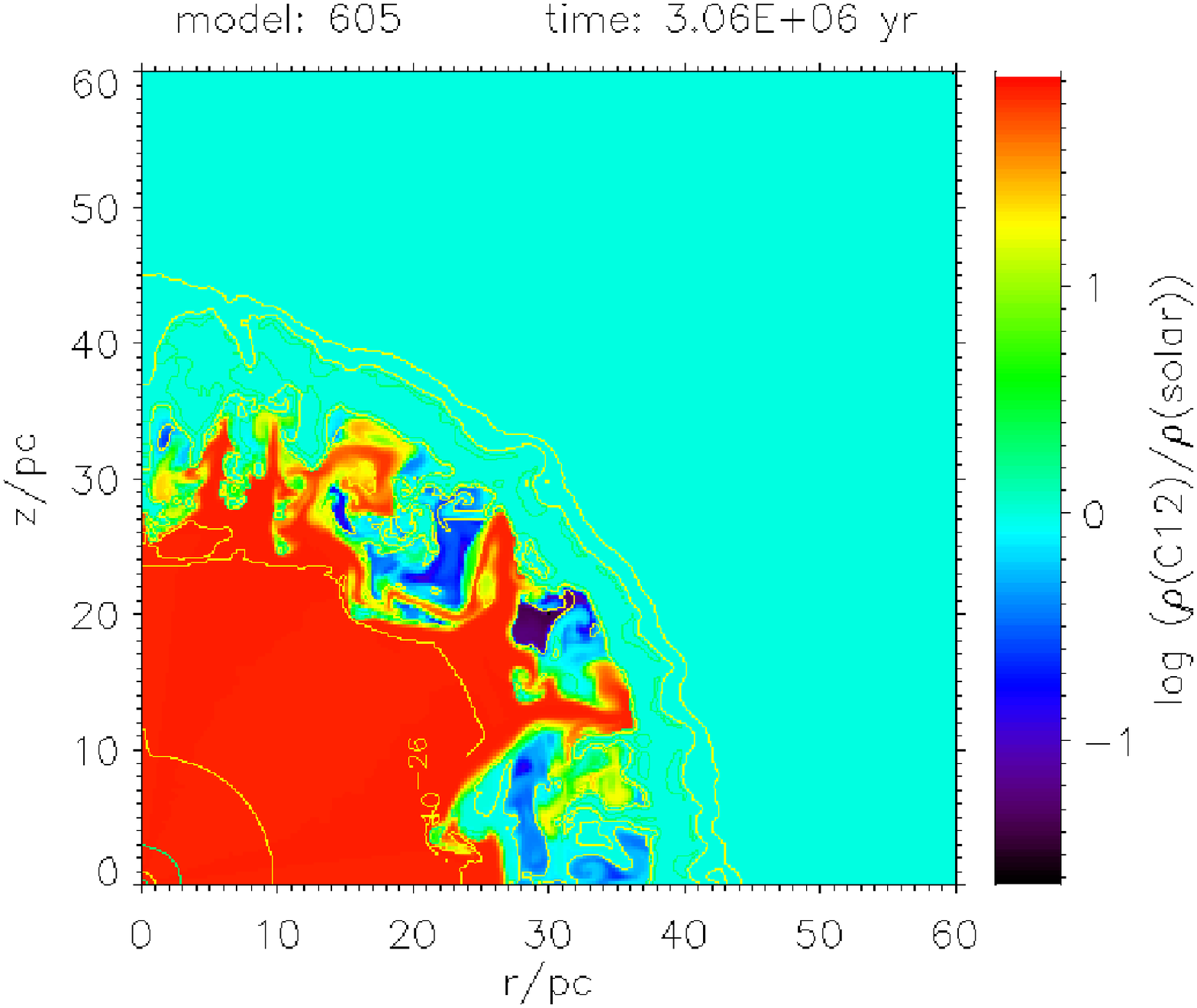}\hspace{-3cm}
\includegraphics{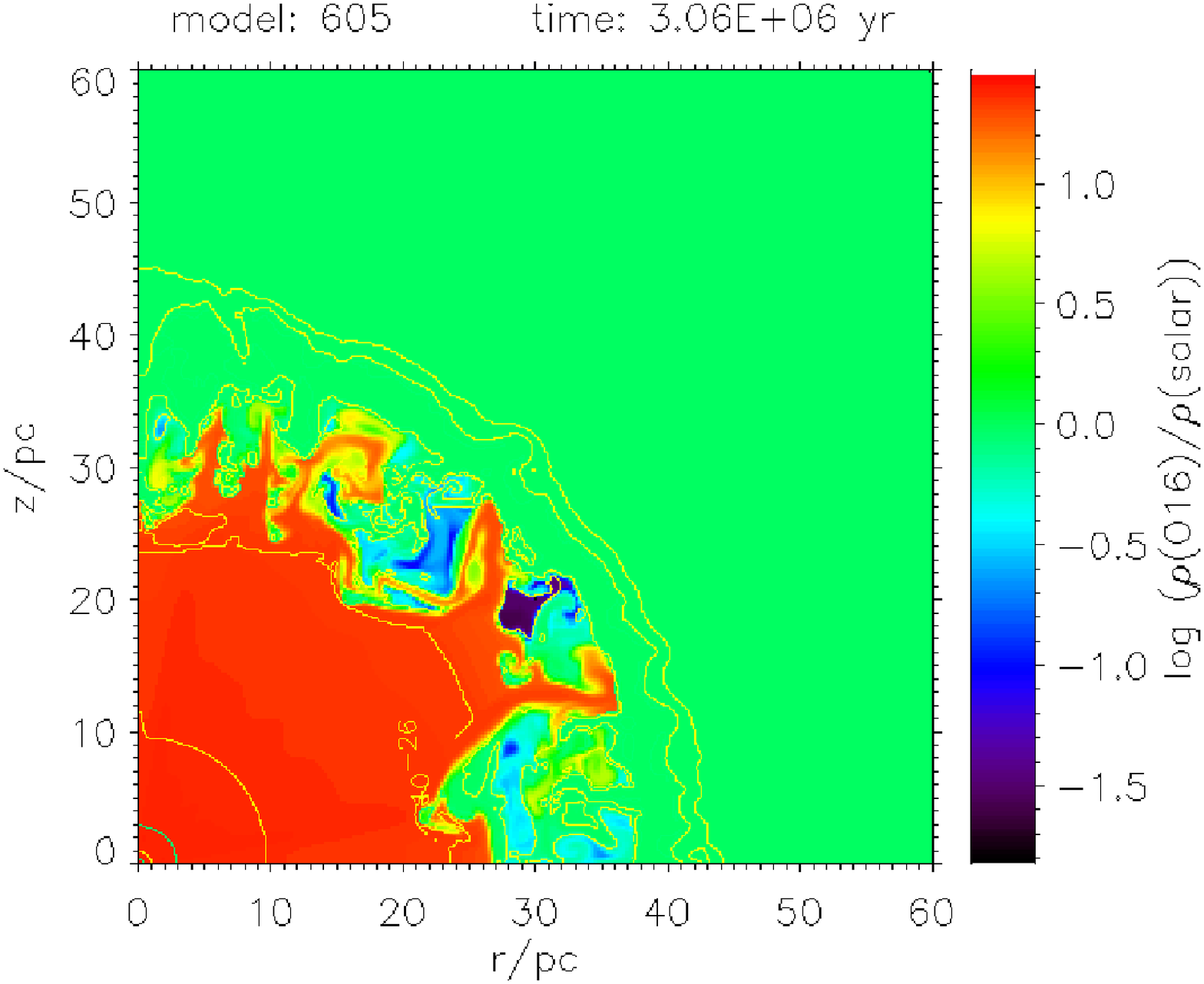}}
\resizebox{18.5cm}{!}{\hspace{-3cm}
\includegraphics{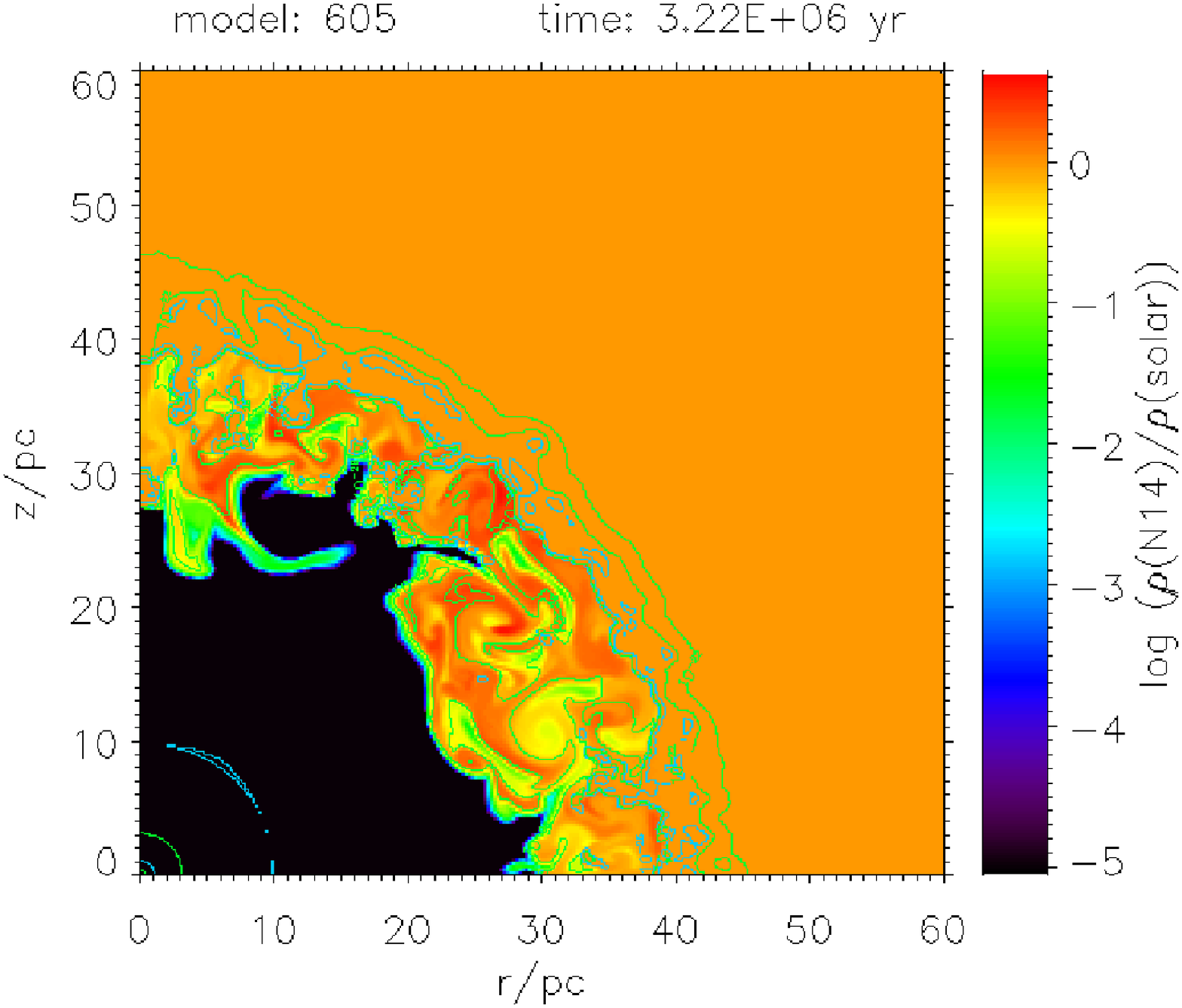}\hspace{-3cm}
\includegraphics{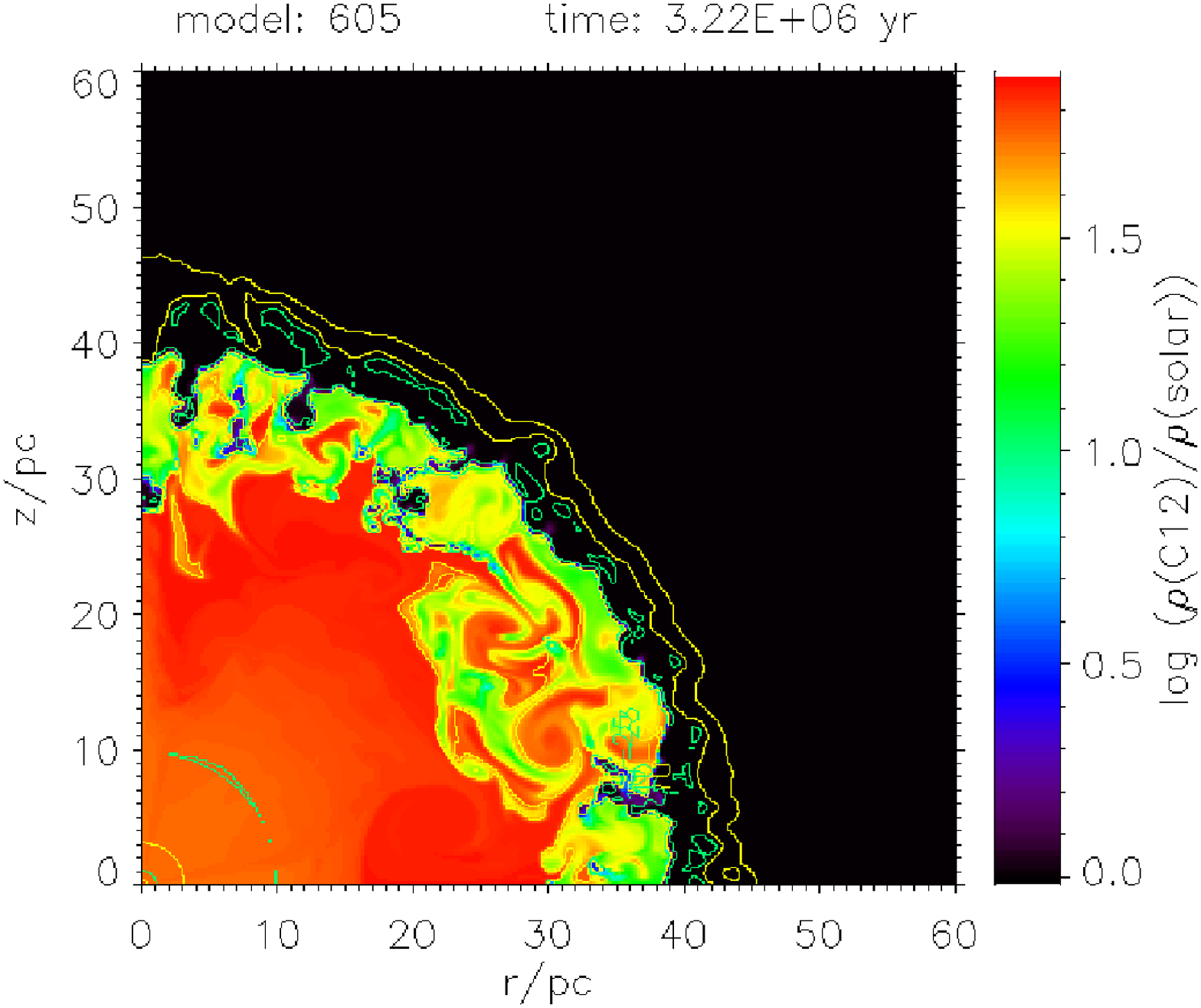}\hspace{-3cm}
\includegraphics{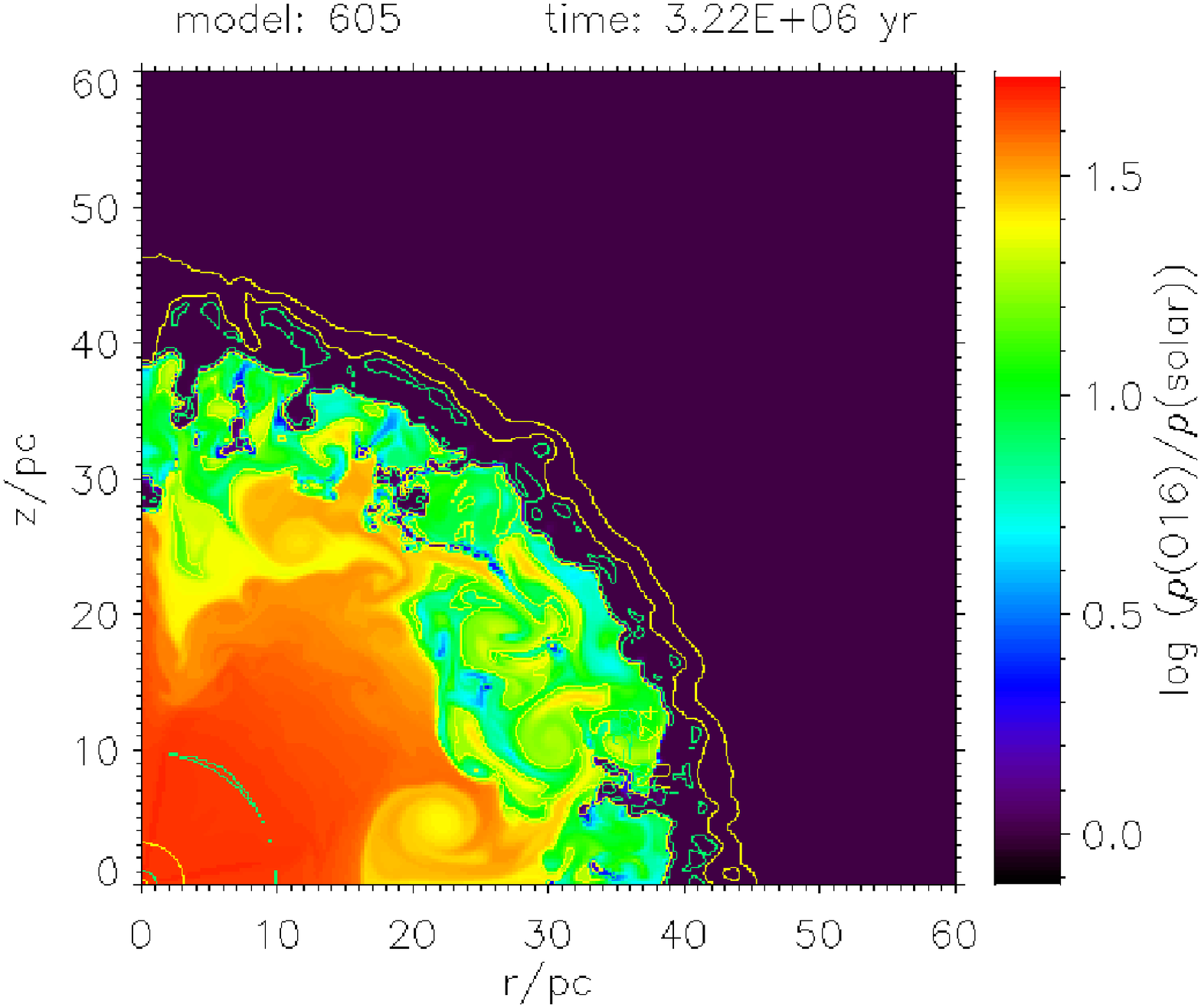}}
      \caption{Concentration of $^{14}{\mbox{N}}$ (left plots),  $^{12}{\mbox{C}}$ (middle), and  $^{16}{\mbox{O}}$ (right plots). Upper row:  $t = 2.99 \mbox{ Myr}$ (top left),  
        near the end of the WN phase, and at $t = 3.06 \mbox{ Myr}$
        (middle and right), after the onset of the WC phase. Lower
        row: At the end of the stellar lifetime ($t = 3.22 \mbox{
          Myr}$).}
      \label{CNO}
\end{figure*}
In the hot gas of the SWB the abundance of $^{14} {\mbox{N}}$ at $t =
2.99 \mbox{ Myr}$ (close to the end of the WN phase,
Fig.~\ref{CNO}, upper left plot) that lasts only for $\approx 0.17
\mbox{ Myr}$, reaches values in the range of $0.1 \lesssim \log \left(^{14}
{\mbox{N}}/\mathrm{solar}\right) \lesssim 0.98$. $^{14} {\mbox{N}}$ is distributed in
the whole SWB except in the H{\sc ii} material enveloped by the SWB.
$^{12}\mbox{C}$ and $^{16}\mbox{O}$ can be found at the same places,
but their concentration is as low as $\log \left(^{12} {\mbox{C}}/\mathrm{solar}\right)
\approx -1.4$ and $\log \left(^{16} {\mbox{O}}/\mathrm{solar} \right) \approx -1.5$.

During its WN phase the star releases 
$0.143 \mbox{ M}_{\sun}$ of $^{14} {\mbox{N}}$,
which is more than half of its total $^{14} {\mbox{N}}$ release, but
almost no extra $^{12} {\mbox{C}}$ or $^{16} {\mbox{O}}$ is supplied.

With the onset of the WC phase at $t = 3.005 \mbox{ Myr}$ the
$^{14}{\mbox{N}}$ release ends and the main enrichment of the SWB with
$^{12}{\mbox{C}}$ and $^{16}{\mbox{O}}$ starts. Due to the beginning
WC phase with its high mass-loss rates and high terminal wind velocity
the previously ejected material in the SWB including the emitted
$^{14}{\mbox{N}}$ is compressed into a shell-like structure.
Therefore, after $t = 3.005 \mbox{ Myr}$ $^{12}\mbox{C}$ and
$^{16}\mbox{O}$ are distributed in the whole SWB and $^{14}\mbox{N}$
can mostly be found in an expanding shell (see lower left plot of
Fig.~\ref{CNO}, at $t = 3.22 \mbox{ Myr}$).  The density contour line
at a radius of $28 \mbox{ pc}$ in the upper middle and upper right
plots of Fig.~\ref{CNO} ($t = 3.06 \mbox{ Myr}$) marks the position of
the reverse shock which borders the free-flowing wind zone.  At $3.06
\mbox{ Myr}$ the radius of the reverse shock has reached its biggest
extent.

During this evolutionary phase (see upper middle and upper right plots
in Fig.~\ref{CNO}, at $3.06 \mbox{ Myr}$) the star ejects large
amounts of $^{12}{\mbox{C}}$ and $^{16}{\mbox{O}}$: The $\log
\left(^{12}\mbox{C}/\mathrm{solar} \right)$ value reaches $\approx
1.86$ inside the highly enriched SWB and the $\log
\left(^{16}\mbox{O}/\mathrm{solar} \right)$ maximum value is $\approx
1.35$.

At the end of its lifetime at $t = 3.22 \mbox{ Myr}$ the $85\mbox{
  M}_{\odot}$ star has supplied $0.28 \mbox{ M}_{\sun}$ of
$^{14}\mbox{N}$, $ 13.76 \mbox{ M}_{\sun}$ of $^{12}\mbox{C}$, and
$11.12 \mbox{ M}_{\sun}$ of $^{16}\mbox{O}$, which are contained in
the combined SWB/H{\sc ii} region.
This situation is shown in the lower plots of Fig.~\ref{CNO}. The
reverse shock is now located at a radius of $9 \mbox{ pc}$, the hot
gas of the SWB fills the area up to a radius of $39 \mbox{ pc}$.
Almost all of the ejected $^{12}\mbox{C}$, $^{14}\mbox{N}$, and
$^{16}\mbox{O}$ is located in the SWB's volume.

In the outer zones of the SWB the highest abundance of $^{12}\mbox{C}$
(see lower middle plot of Fig.~\ref{CNO}) is locally reached with
$\log \left(^{12}\mbox{C}/\mathrm{solar} \right) \approx 1.85$. For
radii $r \lesssim 15 \mbox{ pc}$ the value is $\approx 1.74$.

The $\log \left(^{14}{\mbox{N}}/\mathrm{solar} \right)$ ratio
(Fig.~\ref{CNO}, lower left) reaches $ \approx 0.55$ in the
outer zones of the SWB, where the material ejected during the WN phase
is located. In the inner parts of the SWB only $^{12} {\mbox{C}}$ and
$^{16} {\mbox{O}}$ can be found.

The maximum of the $\log \left(^{16}\mbox{O}/\mathrm{solar} \right)$ value
(Fig.~\ref{CNO}, lower right) of $\approx 1.65$ is reached at
radii $r \lesssim 15 \mbox{ pc}$. Within the outer parts of the SWB the
values range from $0.5$ to $1.4$, highly depending on the location,
since the abundance within the embedded clumps differs from that of
the hot SWB material. 

%
%________________________________________________________________

\begin{figure}
   \resizebox{\hsize}{!}{\includegraphics{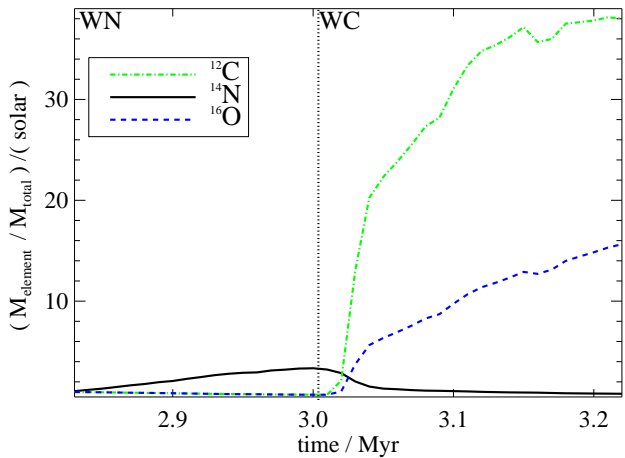}}
   \caption{Time-dependent abundance of ${^{12}\mbox{C}}$, ${^{14}\mbox{N}}$, 
     and ${^{16}\mbox{O}}$ in the hot gas phase. 
            For details see text.}
   \label{C12_N14_O16_hot}
\end{figure}

The temporal evolution of the ${^{12}\mbox{C}}$ abundance in the hot
gas phase is plotted in Fig.~\ref{C12_N14_O16_hot} as dash-dotted
line.  The plot starts at $t=2.83 \mbox{ Myr}$, when the WN phases are
reached. Additionally, the onset of the WC phase at $t=3.005 \mbox{
  Myr}$ is indicated. During the WN phase the stellar parameters
\citepalias{schaller:1992} specify the ${^{12}\mbox{C}}$ abundance of
the WR wind to about $1/14$ solar. Due to the mixing of the WR wind
with the main-sequence wind the averaged mass ratio in the hot gas of
the SWB is only $0.7$ times solar during the WN and early WC phase.
The ${^{12}\mbox{C}}$ enrichment starts with the onset of the WC
phase. After the newly ejected material passed the reverse shock the
${^{12}\mbox{C}}$ abundance in the hot gas reaches $2.1$ times the
solar value at $t = 3.02 \mbox{ Myr}$.  This value increases further
to $38$ times solar at the end of the stellar lifetime.

The $^{14}\mbox{N}$ abundance is depicted in
Fig.~\ref{C12_N14_O16_hot} as solid line. It rises as the enriched
free-floating stellar wind passes the reverse shock, where the
material is heated to several $10^7 \mbox{ K}$. Since in the WC stage
the stellar wind contains no ${^{14}\mbox{N}}$, the
$^{14}\mbox{N}$ abundance in the hot SWB decreases strongly after it
reached its maximum value of $3.4$ times solar at $t = 3.0 \mbox{ Myr}$. 
At the end of the stellar lifetime the $^{14} \mbox{N}$ 
abundance in the hot gas phase is only $0.8 $ times solar.

Because the enrichment with $^{16}\mbox{O}$ starts together with
$^{12}\mbox{C}$ at the onset of the WC phase and due to the high
mass-loss rate in the WC phase the O-abundance in the hot gas phase
rises steeply after the material passed the reverse shock. After $t =
3.02 \mbox{ Myr}$ the $^{16}\mbox{O}$ mass ratio in the hot gas
becomes supersolar and
increases until the end of the stellar lifetime up to $15.7$ times
solar.

%

%________________________________________________________________
%________________________________________________________________

\subsection{Chemical enrichment of the H{\sc ii} region}

During the WR phases of the $85\mbox{M}_{\sun}$ star the SWB expands
into the surrounding H{\sc ii} region without being bordered by a
shock front.  Therefore, the hot SWB gas structures the H{\sc ii}
region and encloses H{\sc ii} gas which is thus compressed into
clumps. This embedded H{\sc ii} material contributes to the warm gas phase
as well as the H{\sc ii} layer around the SWB.

At the end of the stellar lifetime the $^{12}\mbox{C}$ abundance
(lower middle plot in Fig.~\ref{CNO}) in the H{\sc ii} layer
reaches a local maximum of $\log \left(^{12}\mbox{C}/\mathrm{solar}
\right) \approx 0.17$.
It contains a small fraction of $^{14} {\mbox{N}}$, the $\log
\left(^{14} {\mbox{N}}/\mathrm{solar} \right)$ values there are up to
$0.01$ (Fig.~\ref{CNO}, lower left).
The $\log \left(^{16}\mbox{O}/\mathrm{solar} \right)$ values
(Fig.~\ref{CNO}, lower right) range from 0 to $0.03$.

\begin{figure}
   \resizebox{\hsize}{!}{\includegraphics{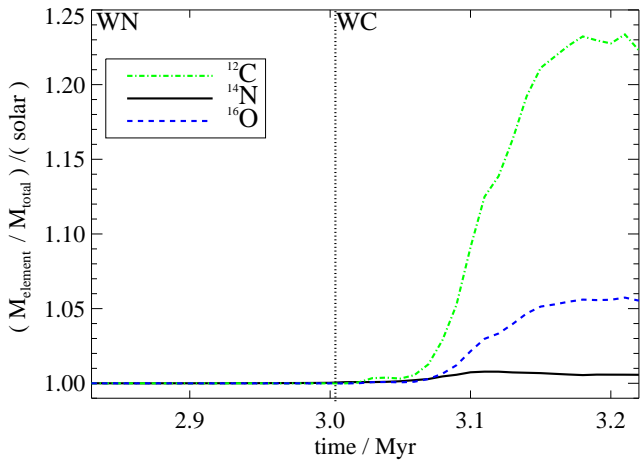}}
   \caption{Same as Fig.~\ref{C12_N14_O16_hot}, for the warm H{\sc ii} gas
      phase. For details see text.}
   \label{C12_N14_O16_warm}
\end{figure}

The temporal evolution of the ${^{12}\mbox{C}}$ abundance is plotted
as dash-dotted line in Fig.~\ref{C12_N14_O16_warm}. As in
Fig.~\ref{C12_N14_O16_hot}, the plot starts with the beginning of the
WN phases, the vertical dotted line indicates the onset of the WC
phase.  In the warm H{\sc ii} gas the averaged $^{12}\mbox{C}$
abundance reaches supersolar values after $t = 3.02 \mbox{ Myr}$, thus
the enriched material needs about $1.5 \times 10^4 \mbox{ yr}$ for its
way from the stellar surface, through the SWB, until it reaches the
H{\sc ii} region and mixes with the warm H{\sc ii} gas. It rises
steeply until reaching a value of $1.21$ times solar at $t = 3.15
\mbox{ Myr}$ and is $1.22$ times solar at $t = 3.22 \mbox{ Myr}$.

The $^{14}\mbox{N}$ mass fraction (solid curve in
Fig.~\ref{C12_N14_O16_warm}) in the warm gas phase rises to supersolar
values after $t = 2.88 \mbox{ Myr}$ when the $^{14}\mbox{N}$ has
passed the SWB and cools down to H{\sc ii} temperatures.  As
discussed for Fig.~\ref{CNO}, after $t = 3.0 \mbox{ Myr}$
the $^{14}\mbox{N}$-enriched material can only be found  in the
shell-like structure in the outer zones of the SWB.  The
${^{14}\mbox{N}}$ abundance in the warm H{\sc ii} gas reaches its
maximum of $1.008 $ times solar at $t = 3.11 \mbox{ Myr}$ and has a
value of $1.006$ times solar at the end of the stellar lifetime.
 
For the warm gas the $^{16}\mbox{O}$ mass ratio is given as dashed
line in Fig.~\ref{C12_N14_O16_warm}. Like $^{12}\mbox{C}$ it reaches
supersolar values after $t = 3.02 \mbox{ Myr}$. Until the end of the
stellar lifetime it increases slightly but reaches only $1.06$ times
the solar value at $t = 3.22 \mbox{ Myr}$.
%________________________________________________________________
%________________________________________________________________

\section{Discussion}
\label{Summary}
 
This chemical analysis of the wind and
radiation-driven H{\sc ii} region around an $85 \mbox{ M}_{\odot}$
star with solar metallicity 
provides first quantitative conclusions to what extent
the C, N, O-enriched WR winds contribute to the observable abundances
of the surrounding H{\sc ii} region.

When comparing the C, N, O abundances in our simulation with observed
values, one has to take into account that the overall lifetime as well
as the WR lifetime of the $85 \mbox{ M}_{\sun}$ star are extremely short
compared to less massive stars. Also, the choice of the stellar
parameters of \citetalias{schaller:1992} influences our results, since
other evolutionary tracks would provide other WR lifetimes as well as
other mass-loss rates (see, e.g., \citetalias{kfhy}).

At the end of the stellar lifetime, the $^{12}\mbox{C}$ abundance in
the warm gas phase amounts to $22.3 \% $ above solar, whereas we find
only $0.6 \%$ supersolar for $^{14}\mbox{N}$, and $5.5 \%$ supersolar
for $^{16}\mbox{O}$. These values measured in the warm gas phase are
the quantities which should be compared with observed H{\sc ii} regions'
emission spectra.

On the other hand, the hot gas of the WR wind bubble is highly
enriched with $^{12}\mbox{C}$ and $^{16}\mbox{O}$ since the onset of
the WC phase, while it was significantly
enriched with $^{14}\mbox{N}$ during the preceeding WN phase.

From these models we conclude that the enrichment of the circumstellar 
environment with  $^{14}\mbox{N}$ and  $^{16}\mbox{O}$ by WR stars 
might be negligible if the $85\mbox{ M}_{\sun}$ star is representative 
for massive stars passing the WR stage.
Only for $^{12}\mbox{C}$ the enrichment of the H{\sc ii} gas is significant.

Since the occurrence of a WR phase is strongly metal dependent the
enrichment with C should also depend on metallicity like e.g. on O.
One should therefore expect that the C gradient of H{\sc ii} regiones
in galactic discs is steeper than that of O. And indeed,
\citet{esteban} found $\Delta \log \left(\mathrm{C}/\mathrm{O} \right)
= - 0.058 \pm 0.018 \mbox{ dex kpc}^{-1}$ for the Galactic disk.

In metal-poor galaxies one would expect less chemical self-enrichment 
because the stellar mass range of the WR occurrence is shrinked and 
shifted towards higher masses and the WR phases are shorter.

\vspace{-1ex}
\bibliography{Hl121}
\bibliographystyle{aa}

\end{document}